\documentclass[prl,aps,floatfix,twocolumn,nopacs,superscriptaddress]{revtex4}

\usepackage[utf8]{inputenc}
\usepackage[american,]{babel}
\usepackage[T1]{fontenc}
\usepackage[pdftex]{graphicx}  
\usepackage{graphicx, xcolor}
\usepackage{dcolumn}
\usepackage{bm}
\usepackage{amsmath,amsthm,amssymb}
\usepackage{hyperref}
\usepackage{xcolor}
\hypersetup{colorlinks,bookmarksopen,bookmarksnumbered,
citecolor=cyan,
linkcolor=cyan,
pdfstartview=false,
urlcolor=cyan}
\usepackage{graphicx}
\usepackage{braket}
\usepackage{wrapfig}
\usepackage{soul}
\usepackage{mathtools}
\usepackage{float}
\usepackage{natbib}

\begin{document}

\title{Energy Transport between Strange Quantum Baths}

\author{Ancel Larzul}
\affiliation{JEIP, USR 3573 CNRS, Coll\`{e}ge de France, PSL Research University, 11 Place Marcelin Berthelot, 75321 Paris Cedex 05, France}
\author{Marco Schir\`o}
\affiliation{JEIP, USR 3573 CNRS, Coll\`{e}ge de France, PSL Research University, 11 Place Marcelin Berthelot, 75321 Paris Cedex 05, France}

\begin{abstract}
Energy transport in quantum many-body systems with well defined quasiparticles has recently attracted interest across different fields, including out of equilibrium conformal field theories, one dimensional quantum lattice models and holographic matter.  Here we study energy transport between \emph{strange quantum baths} without quasiparticles, made by two Sachdev-Ye-Kitaev (SYK) models at temperatures $T_L\neq T_R$ and connected by a Fermi-Liquid system. We obtain an exact expression for the nonequilibrium energy current, valid in the limit of large bath and system size and for any system-bath coupling $V$. We show that the peculiar criticality of the SYK baths has direct consequences on the thermal conductance,
which above a temperature $T^*(V)\sim V^4$ is parametrically enhanced with respect to the linear-$T$ behavior expected in systems with quasiparticles. Interestingly, below $T^*(V)$ the linear thermal conductance behavior is restored, yet transport is not due to quasiparticles. Rather the system gets strongly renormalized by the strange bath and becomes Non-Fermi-Liquid and maximally chaotic. Finally, we discuss the full nonequilibrium energy current and show that its form is compatible with the structure $\mathcal{J}=\Phi(T_L)-\Phi(T_R)$, with $\Phi(T)\sim T^{\gamma}$ and power law crossing over from $\gamma=3/2$ to $\gamma=2$ below $T^*$.
\end{abstract}

\date{\today}
\maketitle

Nonequilibrium heat and energy transport phenomena in strongly interacting quantum matter are attracting interest across condensed matter, atomic physics, statistical mechanics and high-energy physics. From one side new experimental platforms to explore quantum heat transport in mesoscopic systems~\cite{pekola2021colloquium}, ultracold atomic gases~\cite{brantut2013thermoelectric} or strongly correlated quantum materials~\cite{kasahara2018majorana,grissonnanche2019giant} have raised new interest on this topic. In parallel, fresh theoretical understanding on quantum many-body systems far from equilibrium has brought forward new results on energy transport and surprising universalities~\cite{Bernard_2012,bhaseen2015energy}.
A well established paradigm concerns systems in which energy transport is due to well defined quasiparticles such as mesoscopic systems made by ballistic channels, leading to the quantum of thermal conductance recently measured experimentally~\cite{jezouin2013quantum}.  For one dimensional integrable quantum many-body systems, where quasiparticles scatter elastically, several results have been obtained concerning linear response energy transport, both in quantum lattice models~\cite{bertini2020finitetemperature} and in Luttinger Liquids~\cite{kane1996thermal,fazio1998anomalous} as well as on the full nonequilibrium energy current~\cite{karrasch2013nonequilibrium,deluca2014energy,biella2016energy,biella2019ballistic}. The latter was found to display a universal form, predicted by out of equilibrium conformal field theories and related to the Stefan-Boltzmann law~\cite{Bernard_2012,Cardy_2010, Bernard_2016}. Deviations due to irrelevant operators have been also actively discussed~\cite{deluca2014energy,medenjak2021tt}.

The general transport behavior of strongly coupled quantum matter which lacks any quasiparticle is on the other hand much less understood. The  Sachdev-Ye-Kitaev (SYK) model~\cite{sachdev93gapless,parcollet99nonfermi,kitaev} has emerged in recent years as paradigmatic model and building block for strange metals and Non-Fermi Liquids (NFLs)~\cite{song2017strongly,franz2018mimicking,Altland_2019,altland2019sachdev,chowdhury2021sachdevyekitaev}, featuring a peculiar criticality associated to an emergent conformal invariance and which leads to maximal chaos~\cite{Maldacena_2016,kitaev2018softmode}. Understanding transport properties of models in the SYK family and their crossover to more conventional Fermi Liquid (FL) behavior can open new windows in our understanding of exotic phases of matter such as planckian metals~\cite{altland2019sachdev,patel2019theory}.

\begin{figure}[t!]
	\includegraphics[width=\columnwidth]{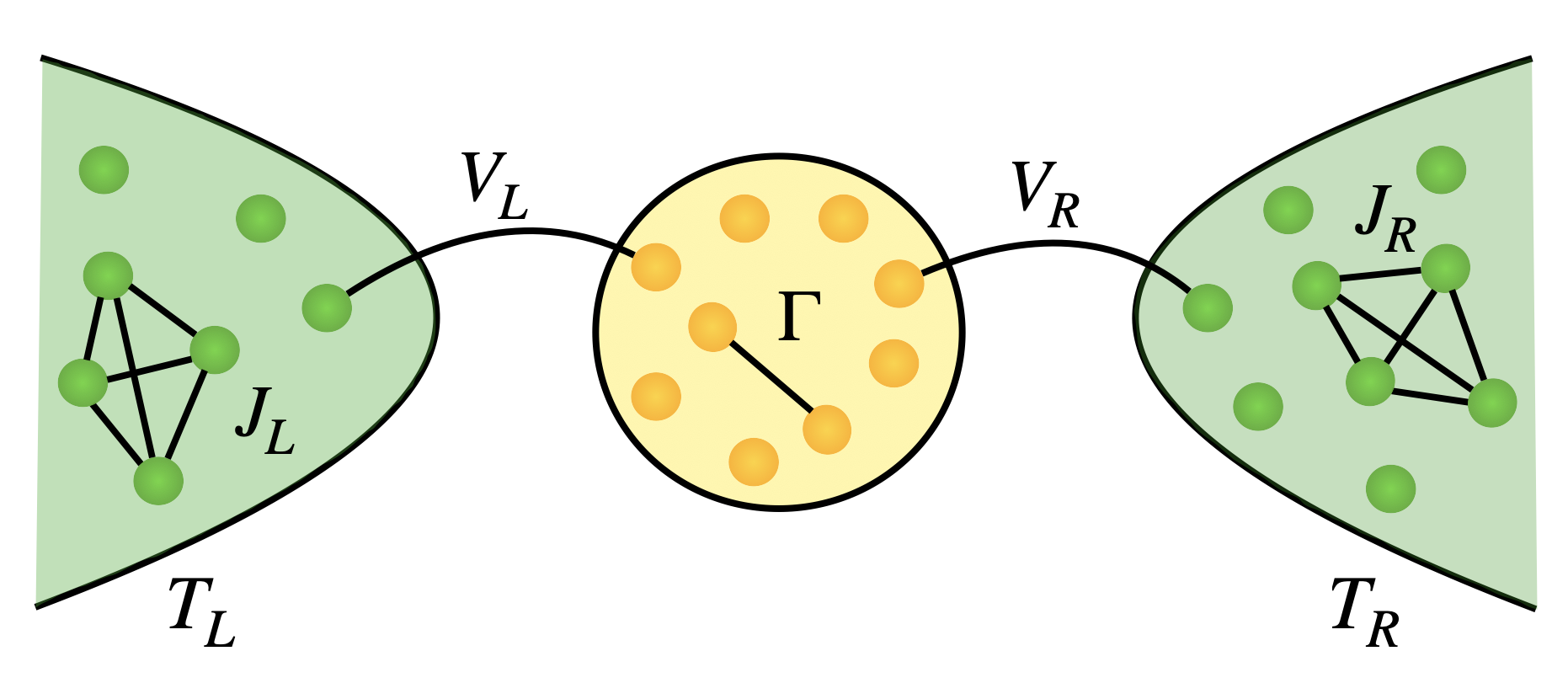}
	\caption{\label{fig:cartoon} Sketch of the setup: two SYK models in equilibrium at different temperatures $T_L\neq T_R$ are connected through a Fermi Liquid dot with random all to all couplings. }
\end{figure}

In this Letter we study the nonequilibrium energy transport between two maximally chaotic reservoirs, described by the SYK model, in equilibrium at different temperatures and connected by tunnel coupling to a FL quantum dot. We note that in the literature the interest has been focused on the study of energy transport and thermal conductivity of SYK-like systems coupled to FL contacts (leads)~\cite{davison2017thermoelectric,kruchkov2020thermoelectric,pavlov2021quantum} or in arrays of SYK dots~\cite{song2017strongly,gu2017local,davison2017thermoelectric,guo2019transport,zanoci2022energy}. Here instead we discuss the role of interactions and maximal chaos in the reservoirs and compare it to the case in which well defined quasiparticles are present both in the system and in the environments. We derive an exact formula for the energy current through this system which takes the form of a generalised Meir-Wingreen formula~\cite{meir1992landauer} for interacting reservoirs. We discuss the linear transport regime and show how the strange metal nature of the bath leaves clear fingerprints in the thermal conductance, which is parametrically enhanced by a weak coupling $V$ to the SYK environment and at low temperature crosses over to a linear temperature scaling. Interestingly we show that the system at low temperature is not a FL, despite the linear thermal conductance, but rather is strongly renormalised by the strange quantum bath, leading to anomalous spectral function and maximal chaos.
Furthermore, we compute the full out of equilibrium energy current and provide evidence that it takes the form $\mathcal{J}= \Phi(T_L)-\Phi(T_R)$, with $\Phi(T)=T^{\gamma}$ and a power-law exponent crossing over from $\gamma=3/2$ to $\gamma=2$, as the temperature goes below a crossover scale $T^*(V)\sim V^4$. 

\emph{SYK Thermal Transport Setup ---}   We consider the transport setup represented in Fig.~(\ref{fig:cartoon}) where two sets of $M$ randomly interacting Majorana fermions $\psi_{a}^{\alpha}$ ($a = 1, \cdots, M$, $\alpha = L, R$) described by the SYK$_4$ model in equilibrium at temperatures $T_L,T_R$, are suddenly connected by an island made of $N$ non-interacting Majorana fermions $\chi_{i}$ ($i = 1, \cdots, N$) with random hoppings, described by the SYK$_2$ model. The total Hamiltonian is
\begin{equation}\label{eq:H_tot}
\begin{split}
    H &= \sum_{\alpha=L,R} H_4^{\alpha} + H^S+\sum_{\alpha=L,R}  H^{S\alpha} 
\end{split}
\end{equation}
where $H^{\alpha}_{4}$ with $\alpha=L/R$ describes the left/right SYK$_4$ reservoirs with Hamiltonian
\begin{equation}
    H_{4}^{\alpha} = - \frac{1}{4!} \sum_{a,b,c,d = 1}^{M}  J_{a b c d} \, \psi_{a}^{\alpha} \psi_{b}^{\alpha} \psi_{c}^{\alpha} \psi_{d}^{\alpha} 
\end{equation}
$H^S$ describes the island of non-interacting Majorana fermions with SYK$_{2}$ Hamiltonian 
\begin{equation}
    H^{S} =  \frac{i}{2} \sum_{i, j = 1}^{N}  \Gamma_{i j} \, \chi_{i} \chi_{j}
\end{equation}
and the remaining terms describe a linear coupling between reservoir and island
\begin{equation}
    H^{S \alpha} = i \,  \sum_{i = 1 }^{N} \sum_{a = 1}^{M} V_{i a} \, \chi_{i} \psi_{a}^{\alpha}
\end{equation}
The couplings entering the Hamiltonian, $J_{a b c d}, \Gamma_{ij}, V_{ia}$  are all independent Gaussian random variables with zero mean and variance respectively $ \overline{J^{2}_{a b c d}} = \frac{3 ! J^{2}}{M^{3}}$, $\overline{V_{i a}^{2}} = \frac{ V^{2}}{M}$ and  $\overline{\Gamma^{2}_{i j}} = \frac{\Gamma^{2}}{N}$. We consider the two reservoirs to be identical and equally coupled to the system and set $J_{L}=J_R=J$ and $V_L=V_R=V$ in the following. We emphasize that the choice of $H^S$ to be non-interacting is made for the sake of highlighting the transport anomalies due to the interacting and maximally chaotic reservoirs, but can be relaxed as we will discuss later on. Finally, we will compare this transport setting to the more conventional case of FL reservoirs described by the SYK$_2$ model, $H_2^{\alpha} =  \frac{i}{2} \sum_{a,b = 1}^{M}  J_{a b} \, \psi_{a}^{\alpha}\psi_{b}^{\alpha}$, where one expects ballistic energy transport due to quasiparticles. We note that in the literature related models have appeared discussing the effect of coupling one (or multiple) non-interacting bath to the SYK$_4$ model and also studying transport~\cite{banerjee2017solvable,chen2017tunable,zhang2019evaporation,can2019solvable,can2019charge}. 
The model is exactly solvable using Keldysh techniques in the limit $N,M\rightarrow\infty$ at fixed $p\equiv N / M$ (see Supplementary Material~\cite{SM}). Here we will focus on the energy transport, namely on the stationary state current that sets at long times through the two reservoirs when $T_L\neq T_R$. 
 
\emph{A Formula for the Energy Current - } Despite our model is fully interacting, the exact solvablity of the SYK$_4$ model allows us to obtain an exact formula for the energy current flowing from one reservoir to the other. In particular using Keldysh techniques, we can compute the current $\mathcal{J}_{\alpha} = \overline{\dot{E}_{\alpha}}(t) = i\overline{ \langle [ H, H_{\alpha} ] \rangle} (t)$ from the lead $\alpha = L,R$   where $\langle \cdots \rangle$ is the average over the Keldysh action while the overline represents average over all disordered couplings. In the nonequilibrium steady-state the energy current between the two reservoirs  $\mathcal{J} \equiv (\mathcal{J}_{L} - \mathcal{J}_{R}) / 2$ can be written as~\cite{SM} 
\begin{equation}
\mathcal{J} = -\frac{ N  V^{2}}{2}  \int \frac{d \omega}{2 \pi} \, \omega \, 
\left(G_L^<(\omega)-G_R^<(\omega)\right)G_{S}^{>}(\omega)
\label{eq:current}
\end{equation}
where $G^{<}_{\alpha}(\omega)$ and $G^{>}_{S}(\omega)$ are, respectively, the lesser and greater components of the Green's function for fermions of the bath $\alpha$ and of the system~\cite{SM}.

Several remarks are in order concerning Eq.~(\ref{eq:current}), which is one of our main result. First, the expression for the energy current in Eq.(\ref{eq:current}) is an exact result in the large $N,M$ limit, at fixed ratio $p=N/M$, and it is non-perturbative in the system-bath coupling $V$. In this respect the Green's functions entering the energy current $\mathcal{J}$ are those fully renormalized by the system-bath interaction~\cite{SM}. 
The structure of Eq.~(\ref{eq:current}) is reminiscent of the  Meir-Wingreen formula usually describing transport between two non-interacting reservoirs connected by an interacting intermediate region~\cite{meir1992landauer,pekola2021colloquium}. This analogy becomes more transparent in the limit $p\ll1$, corresponding to a bath which is parametrically larger than the system, as we are going to consider here. 
With respect to the Meir-Wingreen formula, our result can account for fully interacting and maximally chaotic reservoirs, linearly coupled to a central system. Furthermore we can show that in fact Eq.~(\ref{eq:current}) holds with much more generality, being valid both when reservoirs and systems are non-interacting Majorana fermions with SYK$_2$ Hamiltonian and in presence of random $q_B-$body interactions of the SYK type in both system and reservoirs~\cite{SM}. 
 
\begin{figure}[t!]
	\includegraphics[width=\columnwidth]{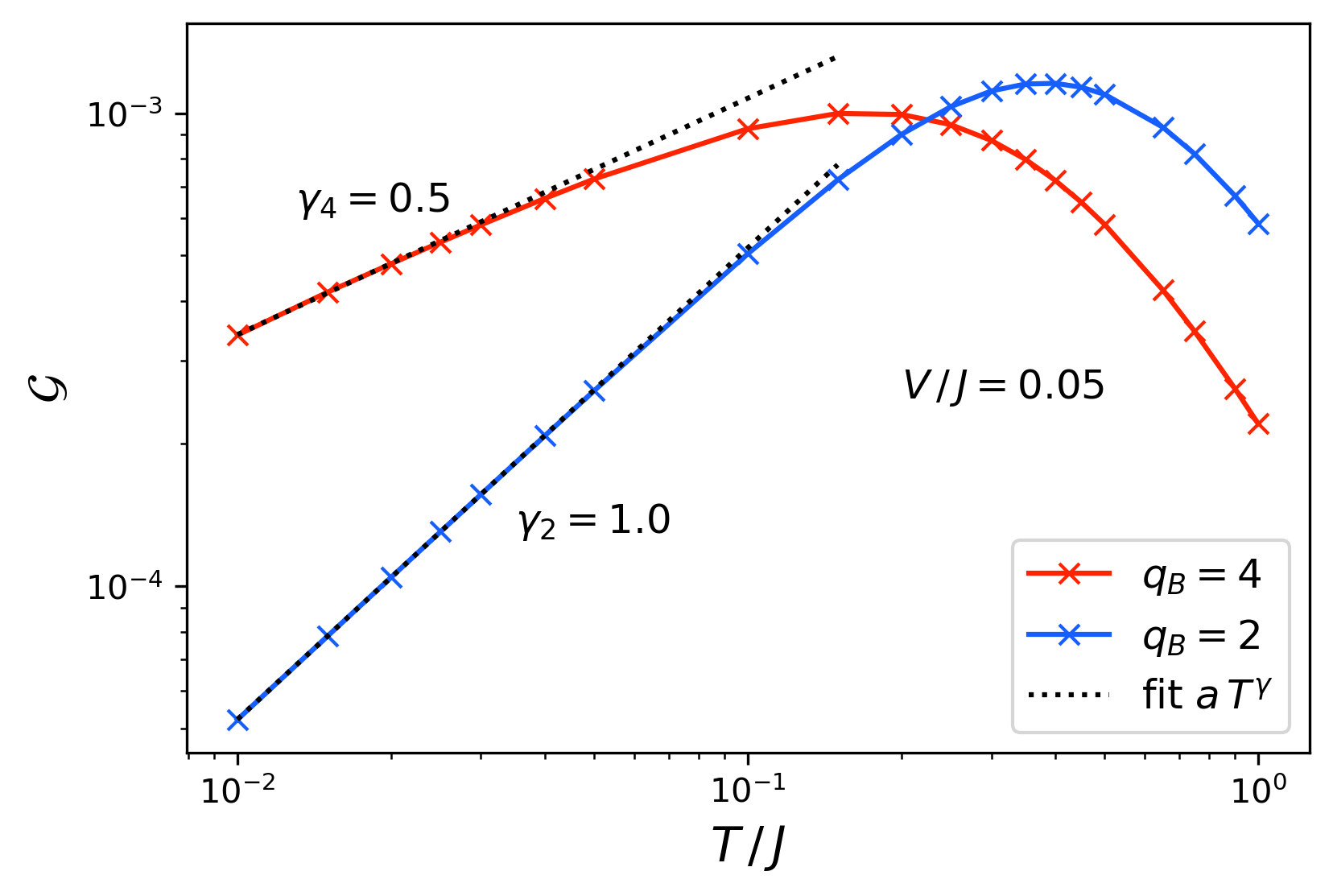}
	\includegraphics[width=\columnwidth]{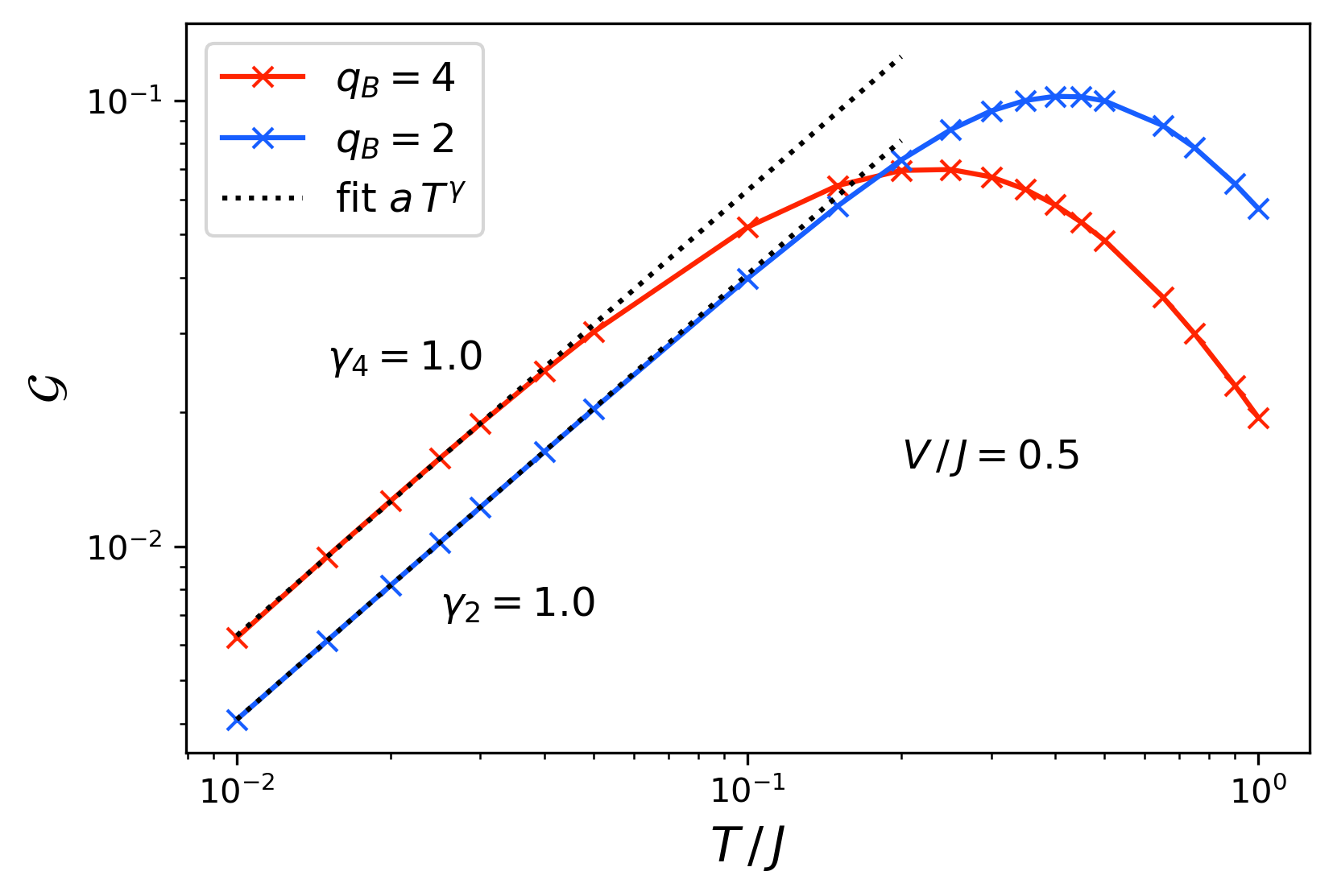}	
		\caption{Linear Thermal conductance $\mathcal{G}(T)$ at weak (top panel) and strong (bottom panel) system-bath coupling, for both conventional SYK$_2$ corresponding to $q_B=2$ and \emph{strange quantum bath} corresponding to $q_B=4$.  }
	\label{fig:conductance}
\end{figure}

\emph{Linear Energy Transport through Strange Quantum Baths - } We start our discussion of energy transport from the linear regime, corresponding to two temperatures differing by a small amount $\Delta T\rightarrow0$, i.e. $T_{L,R} = T \pm \Delta T/2 $. In this case from Eq.~(\ref{eq:current}) we can obtain the thermal conductance $ \mathcal{G}(T)\equiv\mathcal{J}/\Delta T$ as
%
\begin{equation}
    \mathcal{G}(T) = \frac{N V^{2}}{2} \int \frac{d \omega}{2 \pi} \, \omega \, A_{S}^{eq}(\omega) f_{eq}(- \omega) \frac{\partial}{\partial T} \Big( A_{B}^{eq}(\omega) f_{eq}(\omega) \Big)
\label{eq:conductance}
\end{equation}
In this expression, all quantities are evaluated at thermal equilibrium with temperature $T$. In particular $A_{B}^{eq},A_{S}^{eq}$ are the equilibrium spectral density of the bath and the system coupled to it, which can be obtained numerically by solving a Dyson equation~\cite{SM}, while $f_{eq}(\omega)$ is the equilibrium Fermi-Dirac distribution. We note that in the limit $p\ll1$ that we consider here the system is strongly renormalised by the bath, which instead is not affected by the feedback of the system since its size is parametrically larger.

In Fig.~\ref{fig:conductance} we plot the thermal conductance $\mathcal{G}(T)$ as a function of temperature, for both the  SYK$_{2}$ and the SYK$_{4}$ baths and for two different values of the system-bath coupling $V$. We first note that for a FL bath such as SYK$_{2}$ the conductance shows a linear scaling with temperature, independently on the value of $V$ (see top and bottom panels, for $q_B=2$). This result is expected for gapless systems with well defined quasiparticles~\cite{pekola2021colloquium} and can be obtained from Eq.~(\ref{eq:conductance}) by considering that in the low-energy limit $\omega, T \ll J$ the system spectral density becomes flat $A_{S}^{eq}(\omega) \simeq 2 / \Tilde{\Gamma}$ but with a modified coupling constant $\Tilde{\Gamma}$~\cite{SM}, so that we obtain after re-introducing the physical dimensions and writing explicitly $\hbar$ and $k_{B}$, 
\begin{equation}
    \mathcal{G}(T) =  N \, \frac{V^{2}}{\Tilde{\Gamma} J} \, \frac{\pi ^{2} k_{B}^{2}}{3 h} T = N \, \frac{V^{2}}{\Tilde{\Gamma} J} \, \mathcal{G}_{Q}\qquad
\left(q_B=2\right)
\label{eq:cond2}
\end{equation}
where $\mathcal{G}_{Q} =  \frac{\pi ^{2} k_{B}^{2}}{3 h} T $ is the quantum of thermal conductance which corresponds to ballistic transport i.e a probability of transmission across the channel equal to $1$. Thus we can interpret the factor $V^{2} / ( \Tilde{\Gamma J})$ as a typical probability of transmission of an energy carrier from the left reservoir to the right reservoir. 

We now turn to the interacting SYK$_{4}$ reservoirs for which, on the contrary, the thermal conductance $\mathcal{G}(T)$  shows a non-trivial dependence on the system-bath coupling $V$. In particular, as we see in Fig.~\ref{fig:conductance} (top panel, for $q_B=4$), for weak coupling $V/J=0.05$ and low-to-intermediate temperatures the thermal conductance shows a $\sqrt{T}$ scaling, i.e. a \emph{strange quantum bath} leads to an enhanced thermal conductance with respect to a non-interacting  bath of quasiparticles. 

We can understand the origin of this effect by considering the structure of bath and system spectral functions in the  low-frequency conformal limit $\omega, T \ll J$. In fact we know that for the SYK$_{4}$ baths the spectral density is peaked at the origin and is given by the expression
\begin{equation}\label{eq:A_B}
    A_{B}^{eq}(\omega) = 2 \Big( \frac{\pi}{J^{2}}  \Big)^{1/4} \frac{1}{\sqrt{2 \pi T}} \textrm{Re} \, \bigg( \frac{ \Gamma \big( \frac{1}{4} - \frac{i \omega}{2 \pi T} \big) }{ \Gamma \big( \frac{3}{4} - \frac{i \omega}{2 \pi T} \big)} \bigg) 
\end{equation}
where $\Gamma(z)$ is the gamma function and we note that the bath shows quantum critical scaling $\omega/T$. When the tunnel coupling $V$ is weak the spectral density of the system remains close to the isolated one, at least for not too low temperatures, and the Dyson equation can be solved perturbatively in powers of $V^{2} / \Gamma^{2}$. As the energy current has an overall $V^{2}$ factor we can keep the spectral density of the system at order $0$ in $V^{2} / \Gamma^{2}$, so $A_{S}^{eq} \simeq 2 / \Gamma $. Plugging this ansatz and Eq.~(\ref{eq:A_B}) in the expression for the conductance we get up to a numerical prefactor~\cite{SM}
\begin{equation}\label{eq:G4_smallV}
    \mathcal{G}(T) \sim N\frac{V^{2}}{\Gamma \sqrt{J}} \sqrt{T}\qquad
\left(q_B=4\right)
\end{equation}
Similar scaling have been reported for linear thermal transport of an SYK$_4$ model coupled to FL baths~\cite{kruchkov2020thermoelectric}. We emphasize here that the anomalous temperature scaling is a direct consequence of the strange metal nature of the two reservoirs, whose enhanced density of states leads to an increased thermal conductance as compared to the non-interacting SYK$_{2}$ case, i.e. $  \frac{\mathcal{G}_{4}}{\mathcal{G}_{2}} \sim \sqrt{\frac{J}{T}} \gg 1$. 

\begin{figure}[t!]
	\includegraphics[width=\columnwidth]{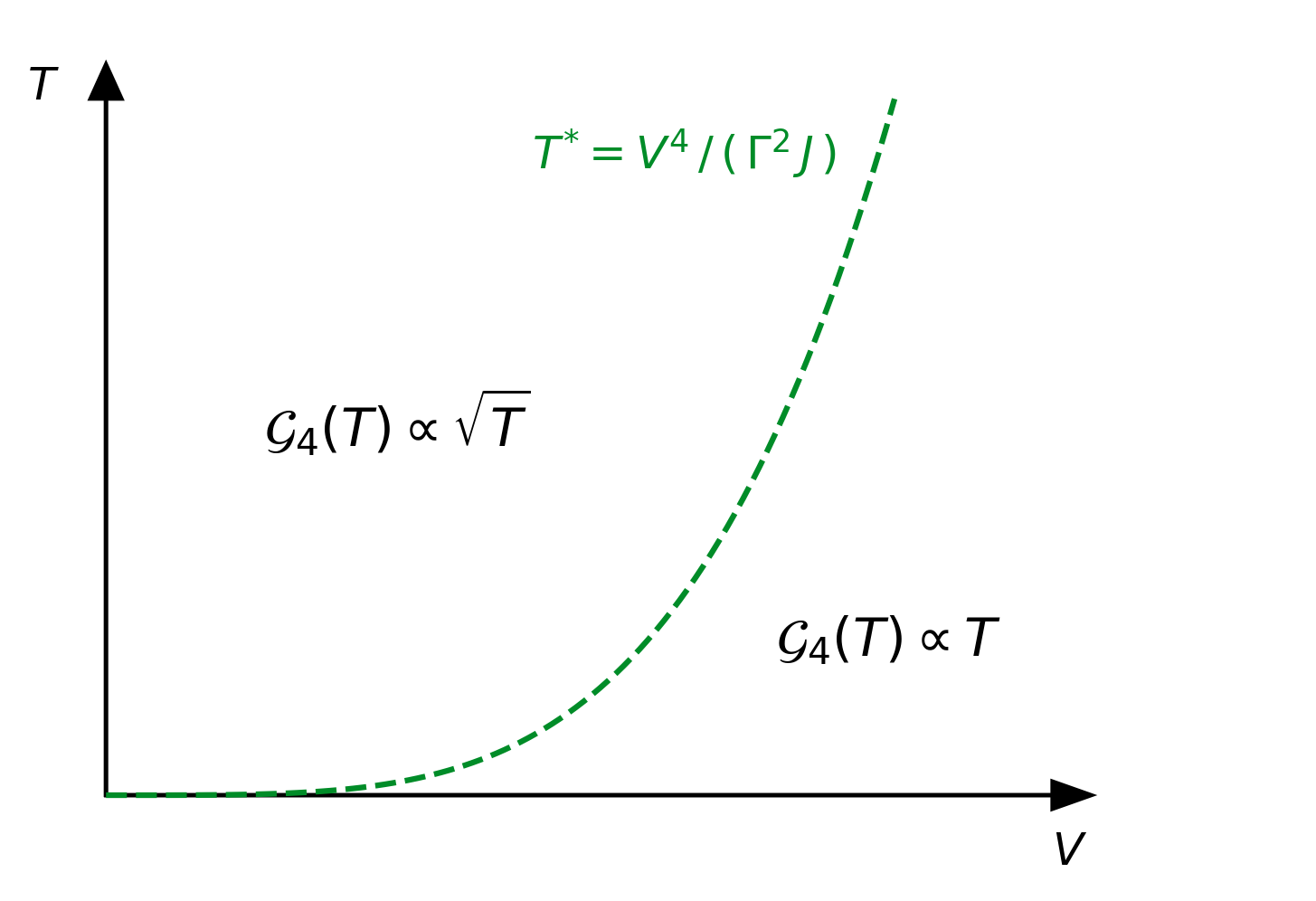}
	\caption{Transport  phase diagram as a function of temperature $T$ and system-bath coupling $V$. Above the crossover temperature $T^*(V)=V^4/ ( \Gamma^2 J )$ the thermal conductance $\mathcal{G}(T)$ is enhanced for the SYK$_4$ bath through the $\sqrt{T}$ scaling, while at low temperatures or strong-coupling the linear behavior is restored.}
	\label{fig:phasediag}
\end{figure}
The parametrically large enhancement in the thermal conductance does not however survive up to strong system-bath couplings, as we see in Fig.~(\ref{fig:conductance}) (bottom panel, for $q_B=4$) where for $V/J=0.5$  the thermal conductance crosses over to a linear temperature scaling. To see how this comes about we note that for larger couplings $V$ the spectral function of the system becomes dressed by the strange metal bath and develops
a dip at small frequency and in particular at zero temperature it scales like $\propto \sqrt{\omega}$, a behavior reminiscent of the zero-bias anomaly in one dimensional disordered interacting conductors~\cite{ALTSHULER19931033,matveev93coulomb,mischenko2001zero}. In the limit $V \gg V^*(T)\equiv ( \Gamma^{2} J T)^{1 / 4}$, which also defines a low-temperature scale $T^*(V)$,
we can obtain an analytic expression for the system spectral function which reads~\cite{SM}
\begin{equation}
    A_{S}^{eq}(\omega) = \frac{1}{V^{2}} \Big( \frac{J^{2}}{\pi}  \Big)^{1/4} \sqrt{2 \pi T} \, \textrm{Re} \bigg( \frac{ \Gamma \big( \frac{3}{4} - \frac{i \omega}{2 \pi T} \big) }{ \Gamma \big( \frac{1}{4} - \frac{i \omega}{2 \pi T} \big)} \bigg)
\end{equation}
The suppressed spectral density of the system renormalised by our strange quantum bath leads to a suppression of thermal conductance and restoring a linear temperature scaling,  as it would be for the FL leads
\begin{equation}
    \mathcal{G}(T) \sim N T\sim N\mathcal{G}_{Q}(T)
    \qquad\left(q_B=4\right)
\label{eq:cond4_a}
\end{equation}
 This similarity is however only superficial, as energy transport in this regime is not due to quasiparticles. The linear $T$-scaling arises in fact from a subtle cancellation between the enhanced spectral density of the SYK$_{4}$ baths and the suppressed density of state of the system and reminds other situations in which violation of a FL scaling leads to a thermal conductance linear in temperature~\cite{vandalum2020wiedemann}. To further appreciate the physics behind this result it is interesting to comment on the relation between transport and chaos in our system. In fact the calculation of the out of time-order correlators for the model in Eq.~\ref{eq:H_tot} in the large $N,M$ limit
  shows~\cite{banerjee2017solvable} that below $T^*(V)$, when the system is dressed by the stange bath and the thermal conductance is linear, the Liapunov exponent saturates the bound on chaos.  In other words, the phase at strong coupling and low-temperature provides an example where transport is suppressed by the strange bath while chaos is enhanced, pointing towards different mechanisms controlling these two processes.

We summarize the linear transport regime of our model in Fig.~(\ref{fig:phasediag}). We note that the crossover scale $T^*(V)$  is strongly  dependent on system-bath coupling and controls also the regime of validity of Eq.~(\ref{eq:G4_smallV}), setting a low-temperature scale below which the $\sqrt{T}$ scaling crosses over to the linear one. Yet at weak coupling $V$ this scale is parametrically small (corresponding to a temperature $T^*\sim 10^{-5}$ for the parameters in Fig.~(\ref{eq:G4_smallV})) leaving a broad range of temperatures where the enhanced conductance is visible.

\begin{figure}[t!]
	\includegraphics[width=\columnwidth]{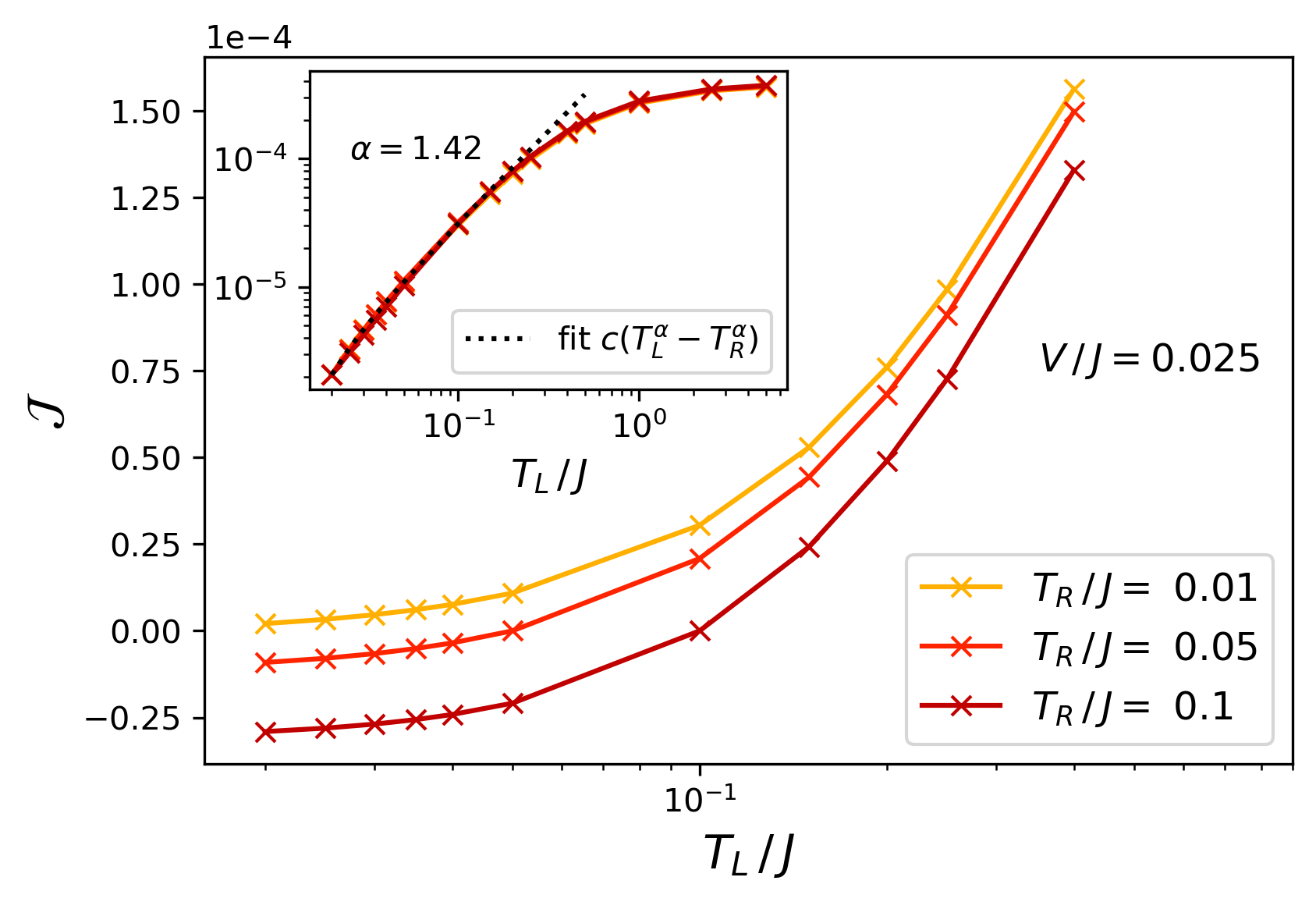}
	\includegraphics[width=\columnwidth]{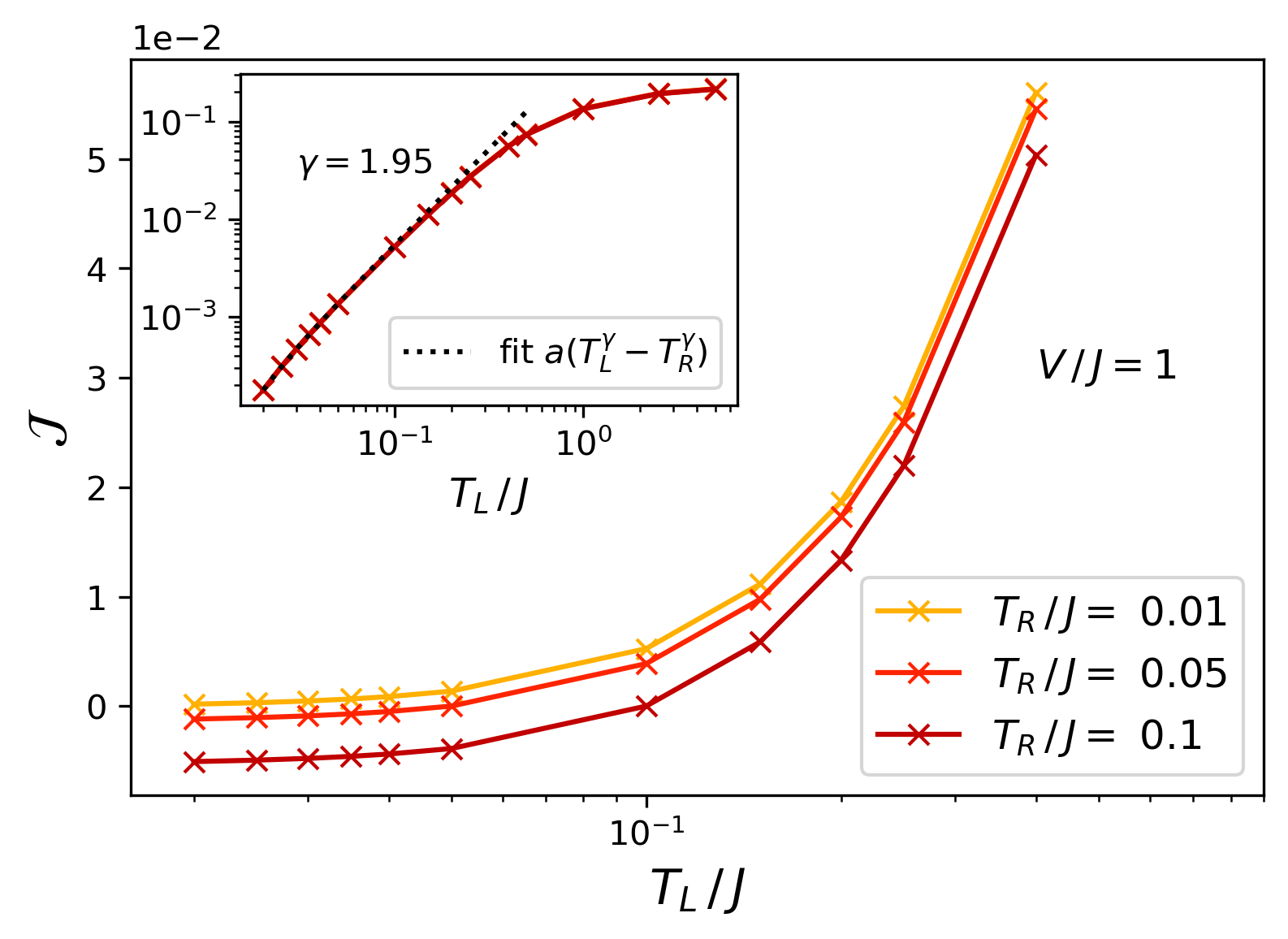}
	\caption{Non-Linear Energy Current $\mathcal{J}$ as a function of $T_L$ and different $T_R$ for weak (top) and strong (bottom) couplings.}
	\label{fig:current}
\end{figure}

\emph{Non-Linear Energy Transport through Strange Quantum Baths - }  Finally, we discuss the full nonequilibrium energy current $\mathcal{J}$ as a function of the two temperatures $T_L,T_R$ and beyond the linear response regime. We first note that for $q_B=2$, corresponding to a non-interacting bath with well defined quasiparticles, the energy current takes the form $\mathcal{J}= \Phi(T_L)-\Phi(T_R)$, with $\Phi(T)\sim T^2$ at low temperatures $T\ll J$, both at weak and strong coupling $V$. This can be seen explictly by using the fact that for a SYK$_2$ bath the spectral function of the system remains temperature independent~\cite{SM}.  This result resembles the one recently obtained in out of equilibrium Conformal Field Theories~\cite{Bernard_2012,Bernard_2016} and implies a Stefan-Boltzmann type of law for the thermal conductance. The situation is richer for SYK$_4$ baths, as we see in Fig.~(\ref{fig:current}) where we plot the energy current $\mathcal{J}$ as a function of $T_L$, for different values of $T_R$ and for weak (top) and strong (bottom) system-bath coupling $V$. In both cases we see that the effect of changing $T_R$ is to induce a rigid shift of the current, suggesting that a functional form of the type $\mathcal{J}= \Phi(T_L)-\Phi(T_R)$ is still compatible with the data despite the fact that, as we know, the spectral function of the system is strongly renormalized by the bath and acquires a rich temperature dependence. As we show in the insets of Fig.~(\ref{fig:current}), for weak system-bath coupling we find $\Phi(T)=T^{3/2}$, a power-law behavior which, while compatible with the thermal conductance discussed earlier, extends for a temperature range well above the linear regime implying a modified Stefan-Boltzmann scaling. For large system-bath coupling on the other hand, or for low enough average temperature $(T_L+T_R)/2$, we see from the inset in the bottom panel that the conventional scaling is recovered $\Phi(T)=T^{2}$, which in this context however does not signal the presence of well-defined quasiparticles.

\emph{Conclusions - } In this work we studied the energy transport between two strange quantum baths, described by the maximally chaotic SYK$_4$ model, coupled through an SYK$_2$ system.  We have obtained an exact formula for the energy current in this setting, which is valid in the large $N,M$ limit at fixed ratio and arbitrary system-bath coupling. We have shown that the quantum-critical nature of the SYK baths has direct consequences on energy transport. The thermal conductance shows a $\sqrt{T}$ scaling above a temperature $T^*$ and crosses over to a linear-$T$ behavior at low temperatures, even though the system becomes Non-Fermi-Liquid and maximal chaotic due to the strange bath. We show that the full nonequilibrium energy current takes the form $\mathcal{J}= \Phi(T_L)-\Phi(T_R)$, with $\Phi(T)\sim T^{\gamma}$ and a power-law exponent $\gamma$ crossing over from $\gamma=3/2$ to $\gamma=2$ below $T^*$. Future directions include considering the charged SYK model to discuss thermoelecricity and the full counting statistics of energy current.

\begin{acknowledgements}
This work was supported by the ANR grant ``NonEQuMat''(ANR-19-CE47-0001). We acknowledge computational resources on the Coll\'ege de France IPH cluster.
\end{acknowledgements}


\widetext
\clearpage

\setcounter{equation}{0}
\setcounter{figure}{0}
\setcounter{table}{0}
\setcounter{page}{1}
\renewcommand{\theequation}{S\arabic{equation}}
\setcounter{figure}{0}
\renewcommand{\thefigure}{S\arabic{figure}}
\renewcommand{\thepage}{S\arabic{page}}
\renewcommand{\thesection}{S\arabic{section}}
\renewcommand{\thetable}{S\arabic{table}}
\makeatletter

\renewcommand{\thesection}{\arabic{section}}
\renewcommand{\thesubsection}{\thesection.\arabic{subsection}}
\renewcommand{\thesubsubsection}{\thesubsection.\arabic{subsubsection}}

\begin{center}
\large{\bf Supplemental Material to `Energy Transport Through a Strange Quantum Bath' \\}
\end{center}

\author{Ancel Larzul}
\affiliation{JEIP, USR 3573 CNRS, Coll\`{e}ge de France, PSL Research University, 11 Place Marcelin Berthelot, 75321 Paris Cedex 05, France}
\author{Marco Schir\'o}
\affiliation{JEIP, USR 3573 CNRS, Coll\`{e}ge de France, PSL Research University, 11 Place Marcelin Berthelot, 75321 Paris Cedex 05, France}

\maketitle

In this Supplemental Material, we provide details on (i) the large $N,M$ Keldysh solution of our model, (ii) the exact formula for the energy current, (iii) the expression for the Green's function of the system in terms of those of the bath, (iv) their expressions in the conformal low-energy limit and (v) the resulting estimates for the thermal conductance.

\section{Keldysh formalism and Schwinger-Dyson equations}

In this section we use Keldysh formalism to derive the exact Schwinger-Dyson equations for the single-particle Green's functions of system and baths in the large $N,M$ limit. We write down the partition function on the closed-time Keldysh contour $Z = \int \mathcal{D}[\chi,\psi{L},\psi^{R}] e^{i S[\chi,\psi^{L},\psi^{R}]}$ with the action $S[\chi, \psi^{L},\psi^{R}]$ 

\begin{equation}
\begin{split}
S[\chi,\psi^{L},\psi^{R}] = \int_{- \infty}^{+ \infty} \sum_{s = \pm} s \Big\{ &\frac{i}{2} \sum_{i,j = 1}^{N} \chi_{i}^{s}(t) \partial_{t} \chi_{i}^{s}(t) + \frac{i}{2} \sum_{i,j = 1}^{M} \sum_{\alpha = L,R} \psi_{a}^{\alpha,s}(t) \partial_{t} \psi_{b}^{\alpha,s}(t)  - \frac{i}{2} \sum_{i,j = 1}^{N}  \Gamma_{i j } \, \chi_{i}^{s} \chi_{i}^{s} \\
&- \frac{i^{\frac{q_{B}}{2}}}{q_{B} !} \sum_{a_{1}, \cdots, a_{q_{B}} = 1}^{M} \sum_{\alpha = L,R} J_{a_{1}, \cdots, a_{q_{B}}} \, \psi_{a_{1}}^{\alpha,s} \cdots \psi_{a_{q_{B}}}^{\alpha,s} - i \, \theta(t) \sum_{i = 1}^{N} \sum_{a = 1}^{M} \sum_{\alpha = L,R} V_{i a} \, \chi_{i}^{s} \psi_{a}^{\alpha,s} \Big\}
\end{split}
\end{equation}

Here $s = \pm$ denotes the upper and lower branches of the closed-time contour and $q_{B} = 2,4$ for the SYK$_{2}$ and SYK$_{4}$ reservoirs respectively. After averaging the partition function over the disorder we can rewrite the action in terms of the bilocal fields

\begin{align}
    G^{s s^{\prime}}_{S}(t_{1},t_{2}) &= - \frac{i}{N} \sum_{i = 1}^{N} \langle \chi_{i}^{s}(t_{1}) \chi_{i}^{s^{\prime}}(t_{2})\rangle = 
    \begin{pmatrix}
    G^{T}_{S}(t,t^{\prime}) & G^{<}_{S}(t,t^{\prime}) \\
    G^{>}_{S}(t,t^{\prime}) & G^{\Tilde{T}}_{S}(t,t^{\prime})
    \end{pmatrix}_{s s^{\prime}}\\
    G^{s s^{\prime}}_{\alpha}(t_{1},t_{2}) &= - \frac{i}{M} \sum_{a = 1}^{M} \langle \psi_{a}^{\alpha,s}(t_{1}) \psi_{a}^{\alpha,s^{\prime}}(t_{2})\rangle
    =
    \begin{pmatrix}
    G^{T}_{\alpha}(t,t^{\prime}) & G^{<}_{\alpha}(t,t^{\prime}) \\
    G^{>}_{\alpha}(t,t^{\prime}) & G^{\Tilde{T}}_{\alpha}(t,t^{\prime})
    \end{pmatrix}_{s s^{\prime}} \qquad \alpha = L,R
\end{align}
which describe the single-particle Green's functions of Majorana fields for the system and the bath respectively, and with the corresponding Lagrange multipliers

\begin{equation}
    \Sigma^{s s^{\prime}}_{S}(t_{1},t_{2}) =
    \begin{pmatrix}
    \Sigma^{T}_{S}(t,t^{\prime}) & - \Sigma^{<}_{S}(t,t^{\prime}) \\
     - \Sigma^{>}_{S}(t,t^{\prime}) & \Sigma^{\Tilde{T}}_{S}(t,t^{\prime})
    \end{pmatrix}_{s s^{\prime}},
    \qquad 
    \Sigma^{s s^{\prime}}_{\alpha}(t_{1},t_{2}) =
    \begin{pmatrix}
    \Sigma^{T}_{\alpha}(t,t^{\prime}) & - \Sigma^{<}_{\alpha}(t,t^{\prime}) \\
     - \Sigma^{>}_{\alpha}(t,t^{\prime}) & \Sigma^{\Tilde{T}}_{\alpha}(t,t^{\prime})
    \end{pmatrix}_{s s^{\prime}}
    \qquad \alpha = L, R
\end{equation}

After integrating over the fermions $\chi$, $\psi^{L}$ and $\psi^{R}$ we get an effective action $S_{eff}$ written only in terms of the fields $G$ and $\Sigma$

\begin{equation}
\begin{split}
    S_{eff}[G,\Sigma] &= - i \frac{N}{2} \textrm{Tr} \log \big[ - i \, \hat{G}_{0,S}^{-1} + i \hat{\Sigma}_{S} \big] - i \, \frac{M}{2} \textrm{Tr} \log \big[ - i \hat{G}_{0,L}^{-1} + i \hat{\Sigma}_{L} \big] - i \, \frac{M}{2} \textrm{Tr} \log \big[ - i \hat{G}_{0,R}^{-1} + i \hat{\Sigma}_{R} \big] \\
    &+ i \,  \frac{N}{2} \int d t \; d t^{\prime} \; \sum_{s s^{\prime}} s s^{\prime} \, \Big( - \frac{\Gamma^{2}}{2} G^{s s^{\prime}}_{S}(t,t^{\prime})^{2} + G^{s s^{\prime}}_{S}(t,t^{\prime}) \Sigma^{s s^{\prime}}_{S}(t,t^{\prime}) \Big)\\
    &+ i \,  \frac{M}{2} \int d t \; d t^{\prime} \; \sum_{\alpha = L,R} \sum_{s s^{\prime}} s s^{\prime} \, \Big( i^{q_{B}} \frac{J^{2}}{q_{B}} G^{s s^{\prime}}_{\alpha}(t,t^{\prime})^{q_{B}} + G^{s s^{\prime}}_{\alpha}(t,t^{\prime}) \Sigma^{s s^{\prime}}_{\alpha}(t,t^{\prime}) \Big)\\
    &- i \,  \frac{N}{2} \int d t \; d t^{\prime} \; \sum_{\alpha = L,R} \sum_{s s^{\prime}}  s s^{\prime} \, \theta(t) \theta(t^{\prime}) V_{\alpha}^{2}  G^{s s^{\prime}}_{S}(t,t^{\prime}) \, G^{s s^{\prime}}_{\alpha}(t,t^{\prime}) \\
\end{split}
\end{equation}

The saddle-point of the action $S_{eff}$ in the large $N,M$ limit gives us the Schwinger-Dyson equations

\begin{equation}
    \Big[  \hat{G}_{0}^{-1} - \hat{\Sigma}_{S} \Big] \circ \hat{G}_{S} = 1, \qquad \Big[  \hat{G}_{0}^{-1} - \hat{\Sigma}_{L,R} \Big] \circ \hat{G}_{L,R} = 1
    \label{eq:dyson}
\end{equation}

and

\begin{align}
    \Sigma^{s s^{\prime}}_{S}(t,t^{\prime}) &= s s^{\prime} \Gamma^{2} G^{s s^{\prime}}_{S}(t,t^{\prime}) + s s^{\prime} V_{L}^{2}  \, \theta(t) \theta(t^{\prime}) \, G^{s s^{\prime}}_{L}(t,t^{\prime}) + s s^{\prime} V_{R}^{2}  \, \theta(t) \theta(t^{\prime}) \, G^{s s^{\prime}}_{R}(t,t^{\prime})\\
    \Sigma^{s s^{\prime}}_{L,R}(t,t^{\prime}) &= - i^{q_{B}} s s^{\prime} J^{2} G^{s s^{\prime}}_{L,R}(t,t^{\prime})^{q_{B} - 1} + p \,  s s^{\prime} V_{L,R}^{2}  \, \theta(t) \theta(t^{\prime}) \, G^{s s^{\prime}}_{S}(t,t^{\prime})
    \label{eq:sigma_psi}
\end{align}

where $[\hat{G}_{0}^{-1}]^{s s^{\prime}}(t,t^{\prime}) = i s \delta_{s s^{\prime}} \delta(t-t^{\prime}) \partial_{t}$ is the free Majorana Green's function
and we have introduced the ratio $p=N/M$. We note that in general the baths Green's functions are coupled to the system's one due to the term in Eq.~(\ref{eq:sigma_psi}) which describes the feedback of the system on the environment and which vanishes in the limit of infinite bath $p \rightarrow 0$. 
As we are going to discuss in the next section, this feedback is crucial in order to generate a finite contribution to the energy current between the interacting baths.\\  

Rather than working in the $s,s'=\pm$ basis it is convenient to introduce the retarded, advanced and Keldysh Green's functions

\begin{align}
    G^{R}_{S}(t,t^{\prime}) &= \theta(t - t^{\prime}) \big( G^{>}_{S}(t,t^{\prime}) - G^{<}_{S}(t,t^{\prime}) \big) \\
    G^{A}_{S}(t,t^{\prime}) &= \theta(t^{\prime} - t) \big( G^{<}_{S}(t,t^{\prime}) - G^{>}_{S}(t,t^{\prime}) \big) \\
    G^{K}_{S}(t,t^{\prime}) &=  G^{>}_{S}(t,t^{\prime}) + G^{<}_{S}(t,t^{\prime}) \\
\end{align}

and likewise for the self-energies. We can perform a rotation to the retarded, advanced and Keldysh basis by multiplying the Dyson equation \ref{eq:dyson} on the left and on the right by the unitary matrix $ U $

\begin{equation}
    U = \frac{1}{\sqrt{2}}
    \begin{pmatrix}
    1 & 1 \\
    1 & - 1
    \end{pmatrix}
\end{equation}

we get

\begin{equation}
    \begin{pmatrix}
    0 & [G_{0}^{A}]^{-1} - \Sigma^{A} \\
    [G_{0}^{R}]^{-1} - \Sigma^{R} & - \Sigma^{K}
    \end{pmatrix} \circ
    \begin{pmatrix}
    G^{K} & G^{R} \\
    G^{A} & 0 
    \end{pmatrix} = \mathbf{1}
\end{equation}

from which we can read out the three Dyson equations on $G_{S}^{R}$, $G_{S}^{A}$ and $G_{S}^{K}$

\begin{align}
    \Big( [G_{0}^{R}]^{-1} - \Sigma^{R} \Big) \circ G^{R} &= 1\\
    \Big( [G_{0}^{A}]^{-1} - \Sigma^{A} \Big) \circ G^{A} &= 1\\
    \Big( [G_{0}^{R}]^{-1} - \Sigma^{R} \Big) \circ G^{K} &= \Sigma^{K} \circ G^{A}\label{eq:dyson_keldysh}
\end{align}

Finally, it can be shown that the first Dyson equation \ref{eq:dyson} can be recast into a more convenient form known as the Kadanoff-Baym equations

\begin{align}
    i \partial_{t_{1}} G^{>,<}_{S}(t_{1},t_{2}) &= \int_{- \infty}^{+ \infty} d t \, \Big( \Sigma^{R}_{S}(t_{1},t) G^{>,<}_{S}(t,t_{2}) +  \Sigma^{>,<}_{S}(t_{1},t) G^{A}_{S}(t,t_{2}) \Big)  \\
    - i \partial_{t_{2}} G^{>,<}_{S}(t_{1},t_{2}) &= \int_{- \infty}^{+ \infty} d t \, \Big( G^{R}_{S}(t_{1},t) \Sigma^{>,<}_{S}(t,t_{2}) + G^{>,<}_{S}(t_{1},t)\Sigma^{A}_{S}(t,t_{2}) \Big)
    \label{eq:KadBaym}
\end{align}

and likewise for the left and right baths.\\

\section{Energy current}


In this section we derive the expression for the energy current flowing across the two baths given in the main text. To this extent we first evaluate the rate of energy flow across the left bath

\begin{equation}
    \mathcal{J}_{L} \equiv \dot{E}_{L} = \overline{ \frac{d }{d t} \langle H_{L}(t) \rangle }
\end{equation}

We can proceed in two ways, either taking the time derivative first and then average over disorder, or do it in the opposite order. Both ways lead to the same result. We chose to take the disorder average first and compute $E_{L}(t)$. In Keldysh formalism, the expectation value of an operator $\mathcal{O}$ can be obtained by introducing  a generating functional $Z[\eta]$

\begin{equation}
    \langle \mathcal{O}(t) \rangle = \frac{i}{2} \lim_{\eta \to 0} \frac{\delta Z[\eta]}{\delta \eta(t)}
\end{equation}

where $Z[\eta]$ is the partition function defined in the previous section except that we shift the Hamiltonia in the Keldysh action $S[\chi,\psi^{L},\psi^{R}]$ by $H \rightarrow H + \eta(t) \mathcal{O}$ on the upper branch and by $H \rightarrow H - \eta(t) \mathcal{O}$ on the lower branch of the time contour. Averaging over the disorder and following the same steps as in the derivation of the Schwinger-Dyson equation in the previous section we find

\begin{equation}
    E_{L}(t) = -M\, i^{q_B+1}  \frac{J^{2}}{q_B} \int_{- \infty}^{t} d t^{\prime} \Big[ G^{>}_{L}(t,t^{\prime})^{q_B} - G^{<}_{L}(t,t^{\prime})^{q_B} \Big]
\end{equation}
 
where $q_B=2$ for the SYK$_{2}$ baths and $q_B = 4$ for the SYK$_{4}$ bath. Taking the derivative with respect to time we get

\begin{equation}
    \frac{d E_{L}(t)}{d t} = - M\,i^{q_B+1} J^{2} \int_{- \infty}^{t} d t^{\prime} \Big[ G^{>}_{L}(t,t^{\prime})^{q_{B}-1} \partial_{t} G^{>}_{L}(t,t^{\prime}) - G^{<}_{L}(t,t^{\prime})^{q_{B}-1} \partial_{t} G^{<}_{L}(t,t^{\prime}) \Big]
\end{equation}

Then we use the expression of the self-energy of the bath, Eq.~(\ref{eq:sigma_psi}), to get

\begin{equation}
\begin{split}
    \frac{d E_{L}(t)}{d t} &= \,i M \int_{- \infty}^{t} d t^{\prime} \Big[ \Sigma^{>}_{L}(t,t^{\prime}) \partial_{t} G^{>}_{L}(t,t^{\prime}) - \Sigma^{<}_{L}(t,t^{\prime}) \partial_{t} G^{<}_{L}(t,t^{\prime}) \Big] \\
    &-i pM V^{2} \int_{- \infty}^{t} d t^{\prime} \Big[ G^{>}_{S}(t,t^{\prime}) \partial_{t} G^{>}_{L}(t,t^{\prime}) - G^{<}_{S}(t,t^{\prime}) \partial_{t} G^{<}_{L}(t,t^{\prime}) \Big]
\end{split}
\end{equation}

Using the Kadanoff-Baym equation for $G_{L}^{>,<}(t,t^{\prime})$, equivalent to Eq.~(\ref{eq:KadBaym}) with $S\leftrightarrow L$,  to replace $\partial_{t} G_{L}^{>,<}(t,t^{\prime})$  one can show that the first integral is zero.  Thus the time derivative of the energy of the $L-$bath is

\begin{equation}
\frac{d E_{L}(t)}{d t} = - i N V^{2} \int_{- \infty}^{t} d t^{\prime} \Big[ G^{>}_{S}(t,t^{\prime}) \partial_{t} G^{>}_{L}(t,t^{\prime}) - G^{<}_{S}(t,t^{\prime}) \partial_{t} G^{<}_{L}(t,t^{\prime}) \Big]
\end{equation}
We see therefore that the presence of a finite energy flow from the bath is an effect of the feedback term between system and bath encoded in the last term of Eq.~(\ref{eq:sigma_psi}).\\

Now we look at the long time limit, assume the system reaches a non-equilibrium steady state with a finite energy current and therefore  that the two-point functions are time translational invariant $G(t,t^{\prime}) = G(t - t^{\prime})$. Then we can introduce the Fourier transform of the Green's functions

\begin{equation}
    G(\omega) = \int d t e^{i \omega t }G(t), \qquad G(t) = \int \frac{d \omega}{2 \pi} e^{- i \omega t } G(\omega) 
\end{equation}

After some simple manipulations we can rewrite $\dot{E}_{L}(t)$ as a single integral over frequencies

\begin{equation}
\mathcal{J}_{L} = - N \, V^{2} \int \frac{d \omega}{2 \pi} \, \omega \, G_{L}^{<}(\omega) G_{S}^{>}(\omega)
\label{eq:JB}
\end{equation}

from which the expression given in the main text for the net current immediately follows.

As already discussed in the main text, we are interested in the regime $p\ll 1$. We can evaluate the Green's functions of system and bath entering the expression of the energy current at $p=0$. This amounts to disregard the feedback term and consider as self energy of the baths the expression
\begin{equation}
    \Sigma^{s s^{\prime}}_{L,R}(t,t^{\prime}) = - i^{q_{B}} s s^{\prime} J^{2} G^{s s^{\prime}}_{L,R}(t,t^{\prime})^{q_{B} - 1}
\end{equation}
and so we can consider that the $\psi^{L,R}$ fermions are isolated and are not affected by the small system. As an immediate consequence we can safely assume that the left and right baths are in thermal equilibrium and satisfy Fluctuation-Dissipation Theorem at temperature $T_L,T_R$ respectively.

\begin{equation}
    G^{<}_{L}(\omega) =  i A_{L}(\omega) f_{L}(\omega), \qquad G^{<}_{R}(\omega) =  i A_{R}(\omega) f_{R}(\omega)
\end{equation}

where $f_{L}(\omega)$ and $f_{R}(\omega)$ are the Fermi-Dirac distributions of the left and right baths respectively. Thus we finally arrive at the expression 
\begin{equation}
    \mathcal{J} = - \frac{i N  V^{2}}{2}  \int \frac{d \omega}{2 \pi} \, \omega \, \Big(  A_{L}(\omega) f_{L}(\omega) - A_{R}(\omega) f_{R}(\omega) \Big)  G_{S}^{>}(\omega)
\end{equation}

\section{Formal Solution of Dyson Equation for System Green's function}

In this section we show how to get an exact expression of $G^{>}_{S}(\omega)$ in terms of the left and right baths Green's functions. We assume that the system is in a stationary state, possibly nonequilibrium, and that two-point functions are time translational invariant. In particular the greater/lesser self-energy of the system is

\begin{equation}
    \Sigma^{>,<}_{S}(t) = \Gamma^{2}\,  G^{>,<}_{S}(t) + V_{L}^{2}\,  G^{>,<}_{L}(t) + V_{R}^{2}\,  G^{>,<}_{R}(t)
    \label{eq:self_energy_gl}
\end{equation}

$\Sigma_{S}^{>,<}$ is a linear function of the Green's functions, so the retarded self-energy $\Sigma_{S}^{R}(t) = \theta(t) \big( \Sigma_{S}^{>}(t) - \Sigma_{S}^{<}(t) \big)$ takes exactly the same form

\begin{equation}
    \Sigma^{R}_{S}(t) = \Gamma^{2} G^{R}_{S}(t) + V_{L}^{2}  G^{R}_{L}(t) + V_{R}^{2}  G^{R}_{R}(t)
\end{equation}

Taking the Fourier transform of this equation and plugging $\Sigma_{S}^{R}(\omega)$ into the Dyson equation $G_{S}^{R}(\omega)^{-1} = \omega - \Sigma_{S}^{R}(\omega)$ we arrive at the quadratic equation on $G_{S}^{R}(\omega)$

\begin{equation}
\Gamma^{2} G_{S}^{R}(\omega)^{2} - \big[ \omega - V_{L}^{2} G_{L}^{R}(\omega) - V_{R}^{2} G_{R}^{R}(\omega) \big] G_{S}^{R}(\omega) + 1 = 0
\label{eq:dyson_eq}
\end{equation}

The solution to this equation is

\begin{equation}
    G_{S}^{R}(\omega) = \frac{1}{2 \Gamma^{2}} \Big[ \omega - S^{R}(\omega) - \delta(\omega)  \Big]
\end{equation}

where we called $S^{R}(\omega) \equiv V_{L}^{2} G_{L}^{R}(\omega) + V_{R}^{2} G_{R}^{R}(\omega) $ and $\delta(\omega) = x(\omega) + i y(\omega)$ is given by

\begin{equation}
    x = \textrm{sign}\Big( \omega - \textrm{Re}S^{R}(\omega) \Big) \sqrt{\frac{1}{2} \Big( \sqrt{B^{2} + C^{2}} + B \Big)}, \qquad y =  \sqrt{\frac{1}{2} \Big( \sqrt{B^{2} + C^{2}} - B \Big)}
\end{equation}

where

\begin{align}
     B &=  \omega^{2} - 4 \Gamma^{2} - 2 \omega \textrm{Re} S^{R}(\omega) + \textrm{Re}^{2} S^{R}(\omega) - \textrm{Im}^{2} S^{R}(\omega) \\
     C &=  - 2 \omega \textrm{Im} S^{R}(\omega) + 2 \textrm{Re}S^{R}(\omega) \textrm{Im}S^{R}(\omega)
\end{align}

We continue by deriving an expression for the Keldysh Green's function of the system. We write equation \ref{eq:dyson_keldysh} in Fourier space and use equation \ref{eq:self_energy_gl} to get an expression of $\Sigma_{S}^{K}(\omega)$ 

\begin{equation}
    G_{S}^{K}(\omega) = G_{S}^{R}(\omega) \Sigma_{S}^{K}(\omega) G^{A}_{S}(\omega), \qquad \Sigma_{S}^{K}(\omega) = \Gamma^{2} G_{S}^{K}(\omega) + V_{L}^{2} G_{L}^{K}(\omega) + V_{R}^{2} G_{R}^{K}(\omega)
\end{equation}

Combining these two equations we get

\begin{equation}
\begin{split}
    G_{S}^{K}(\omega) &= \frac{G_{S}^{R}(\omega) G_{S}^{A}(\omega) }{1 - \Gamma^{2} G_{S}^{R}(\omega) G_{S}^{A}(\omega) } \Big( V_{L}^{2} G_{L}^{K}(\omega) + V_{R}^{2} G_{R}^{K}(\omega)  \Big)\\
    &= \frac{A_{S}(\omega)}{V_{L}^{2} A_{L}(\omega) + V_{R}^{2} A_{R}(\omega)} \Big( V_{L}^{2} G_{L}^{K}(\omega) + V_{R}^{2} G_{R}^{K}(\omega)  \Big) 
\end{split}
\end{equation}

where in the second line we used the two Dyson equation $G_{S}^{R,A}(\omega)^{-1} = \omega - \Sigma_{S}^{R,A}(\omega)$ to rewrite the first factor. Assuming that the two baths satisfy FDT

\begin{equation}
    G_{L}^{K}(\omega) = - i A_{L}(\omega) \tanh\Big( \frac{\beta_{L} \omega}{2} \Big), \qquad
    G_{R}^{K}(\omega) = - i A_{R}(\omega) \tanh\Big( \frac{\beta_{R} \omega}{2} \Big)
\end{equation}

we arrive at

\begin{equation}
    G_{S}^{K}(\omega) = - i A_{S}(\omega) \, \frac{V_{L}^{2} A_{L}(\omega)  \tanh\Big( \frac{\beta_{L} \omega}{2} \Big) + V_{R}^{2} A_{R}(\omega)  \tanh\Big( \frac{\beta_{R} \omega}{2} \Big) }{V_{L}^{2} A_{L}(\omega) + V_{R}^{2} A_{R}(\omega)}
\end{equation}

From this we finally get for the lesser component $ G_{S}^{>}(\omega) = \big(G_{S}^{K}(\omega)  - i A_{S}(\omega) \big) / 2 $

\begin{equation}
G_{S}^{>}(\omega) = - i A_{S}(\omega) f_{S}(- \omega), \qquad \textrm{with} \qquad f_{S}(\omega) =  \frac{f_{L}(\omega) A_{L}(\omega)  +  f_{R}(\omega)  A_{R}(\omega)  }{ A_{L}(\omega)  + A_{R}(\omega) }
\end{equation}

where we assumed $V_{L} = V_{R} = V$ for simplicity. This form of $G_{S}^{>}(\omega)$ reminds of FDT and indeed in the case where $J_{L} = J_{R} = J$ and $T_{L} = T_{R} = T $, $f_{S}(\omega)$ reduces to the Fermi-Dirac distribution $f_{eq}(\omega)$  at temperature $T$. This suggests to interpret $f_{S}(\omega)$ as the steady-state distribution of the system coupled to the reservoirs. Notice that in the case of the non-interacting SYK$_{2}$ reservoirs, $A_{L}(\omega) = A_{R}(\omega)$ (assuming $J_{L} = J_{R}$) and $f_{S}$ is simply

\begin{equation}
    f_{S}(\omega) = \frac{f_{L}(\omega) + f_{R}(\omega) }{2}, \qquad \qquad q_{B} = 2
\end{equation}


\begin{figure}[t!]
	\includegraphics[width=0.3\columnwidth]{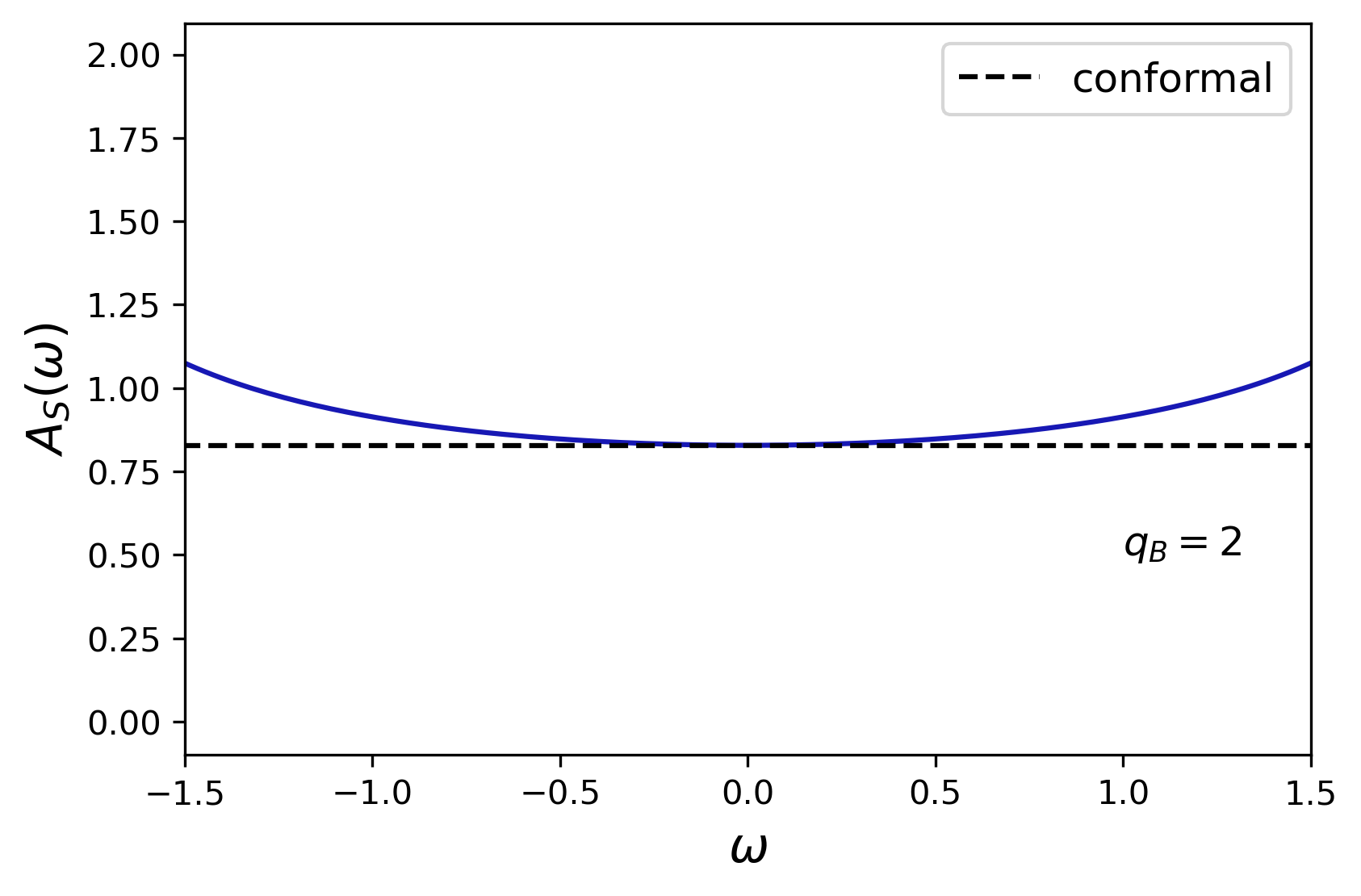}
	\includegraphics[width=0.3\columnwidth]{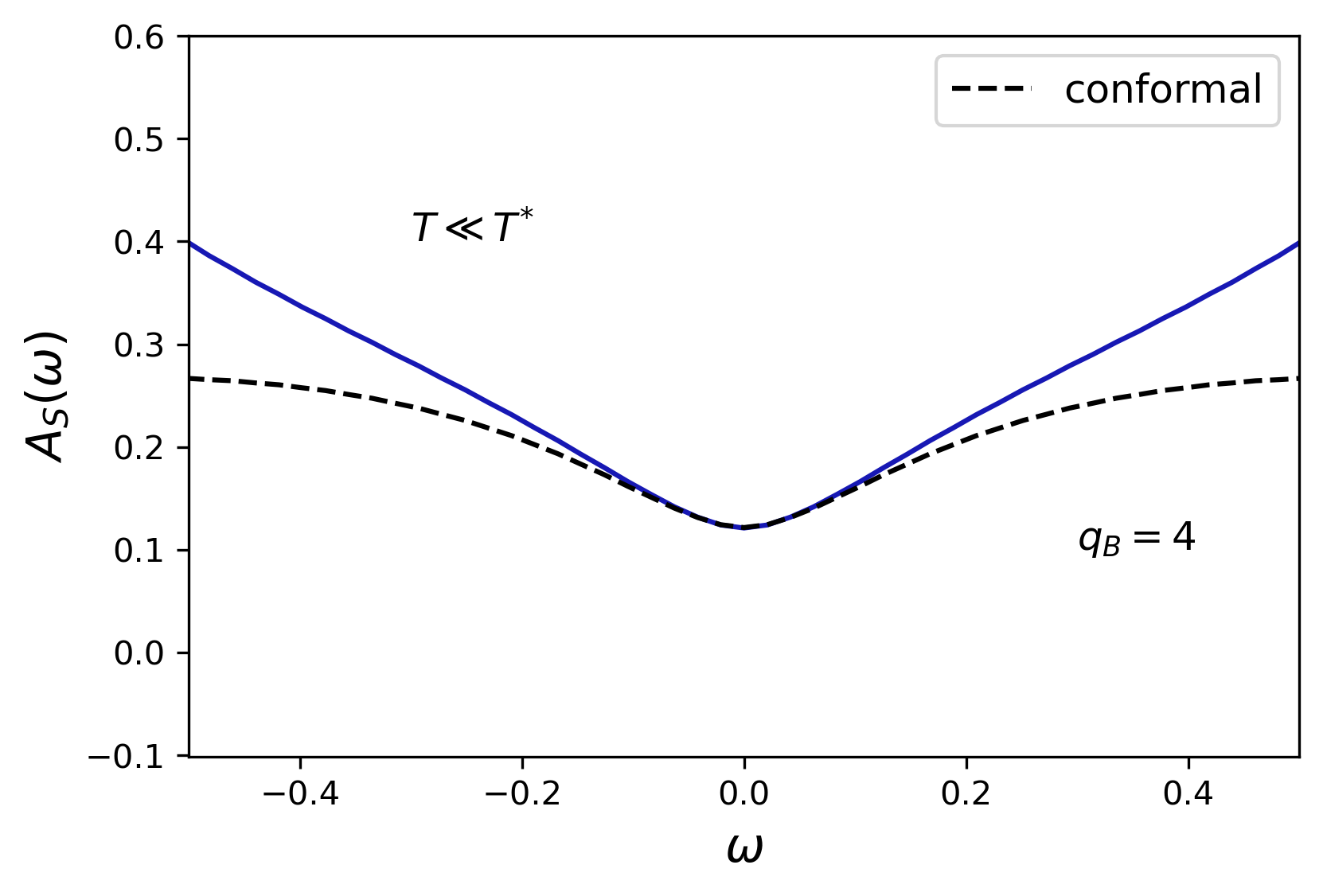}
	\includegraphics[width=0.3\columnwidth]{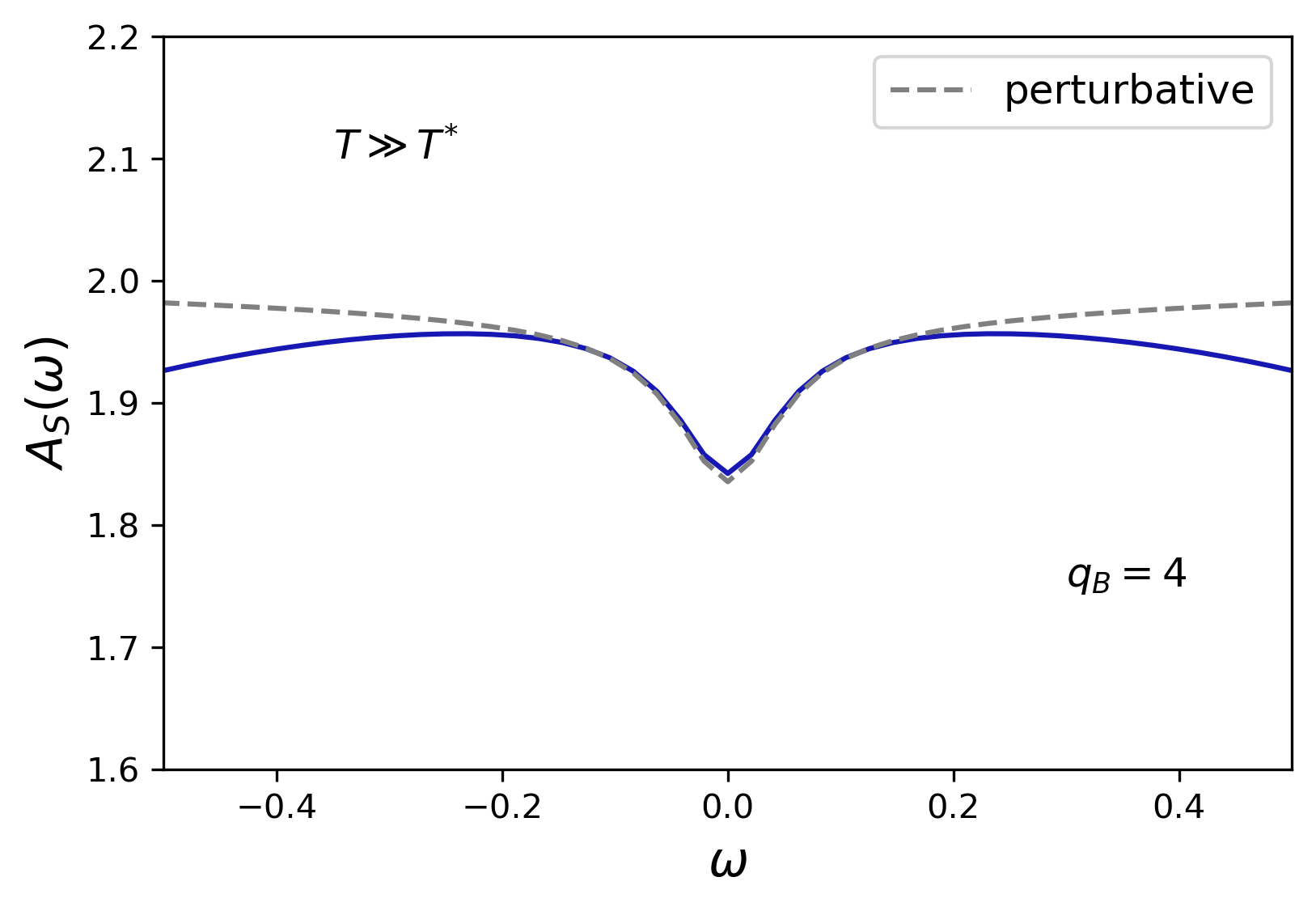}	
	\caption{Spectral Functions of the system renormalized by the bath. For the three plots $\Gamma = J_{L} = J_{R} = 1$, $T_{L} = 0.05$ and $T_{R} = 0.03$ (a) Non-interacting SYK$_{2}$ reservoirs $q_{B} = 2$. The black dashed line is the low energy conformal solution (\ref{eq:G2_conf}). Here $V_{L} = V_{R} = 1$. (b) Strongly interacting SYK$_{4}$ reservoirs for $V_{L} = V_{R} = 1$. In this case, $T^{*} \simeq 1 \gg T_{L}, T_{R}$ and the conformal solution (\ref{eq:G_S4}) correclty reproduces the low energy behaviour (dotted black line). (c) Strongly interacting SYK$_{4}$ reservoirs for $V_{L} = V_{R} = 0.1$. Now $T^{*} \simeq 10^{-4} \ll T_{L}, T_{R} $ and one must use the pertubative expansion (\ref{eq:G_4_p}) to find the low energy behaviour of $A_{s}(\omega)$ (dotted gray line).   } 
	\label{fig:spectralfunction}
\end{figure}

\section{Analytic Expression for System Green's functions in the conformal limit}

In this section we obtain an analytic expression for the system Green's functions in the low-energy conformal limit, using the well known conformal expressions for the  SYK$_2$ and SYK$_4$ baths. We start from the former, i.e. $q_B=2$, and neglet in Eq.~\ref{eq:dyson_eq} the term $\omega$ in the low-energy limit. We also replace $G_{L}^{R}(\omega)$ and $G_{R}^{R}(\omega)$ by their conformal expressions $G_{L}^{R}(\omega) \simeq -i / J_{L}$ and  $G_{R}^{R}(\omega) \simeq -i / J_{R}$ (REF) which yields

\begin{equation}
    \Gamma^{2} G_{S}^{R}(\omega)^{2} - i \Big( \frac{V_{L}^{2}}{J_{L}}  + \frac{V_{R}^{2}}{J_{R}} \Big) G_{S}^{R}(\omega) + 1 = 0
\end{equation}

whose solution is

\begin{equation}
G_{S}^{R}(\omega) = - \frac{i}{\Tilde{\Gamma}}, \qquad \Tilde{\Gamma} = \frac{\Gamma}{\sqrt{1 + \Big( \frac{V_{L}^{2}}{2 \Gamma J_{L}} + \frac{V_{R}^{2}}{2 \Gamma J_{R}} \Big)^{2}} - \Big( \frac{V_{L}^{2}}{2 \Gamma J_{L}} + \frac{V_{R}^{2}}{2 \Gamma J_{R}} \Big) }
\label{eq:G2_conf}
\end{equation}
As we see from this result, the effect of the SYK$_{2}$ baths is just to dress the coupling constant of the SYK$_{2}$ $\chi$ fermions. This is consistent with the fact that with our Hamiltonian $V_{L}$ and $V_{R}$ are marginal perturbations with respect to the SYK$_{2}$ fixed point. Thus the scaling dimension of the $\chi$ fermions remains $ \Delta_{\chi} = 1 / 2$.\\

We now consider the SYK$_4$ baths and, as done before, we neglect the term $\omega$ in the Dyson equation \ref{eq:dyson_eq} . Besides, now with the SYK$_{4}$ baths, $V_{L}$ and $V_{R}$ are relevant perturbations with respect to the SYK$_{2}$ fixed point and the $\chi$ fermions acquire the scaling dimension $\Delta_{\chi} = 3 / 4$. Thus to find the low energy behaviour of $G_{S}^{R}(\omega)$ we can try to neglect also the term $\Gamma^{2} G_{S}^{R}(\omega)$ in the Dyson equation and we get

\begin{equation}
    G_{S}^{R}(\omega) = \frac{-1}{V_{L}^{2} G_{L}^{R}(\omega) + V_{R}^{2} G_{R}^{R}(\omega)}
\label{eq:G_S4}
\end{equation}

We recall that in the conformal limit $G_{L,R}^{R}(\omega)$ are given by

\begin{equation}
G_{L,R}^{R}(\omega) = - i \Big( \frac{\pi}{J_{L,R}^{2}} \Big)^{1/4} \frac{1}{\sqrt{2 \pi T_{L,R}}} \, \frac{\Gamma \big(\frac{1}{4} - i \frac{\omega}{2 \pi T_{L,R}} \big)}{\Gamma \big(\frac{3}{4} - i \frac{\omega}{2 \pi T_{L,R}} \big)}
\label{eq:G_bath}
\end{equation}

This solution for $G_{S}^{R}(\omega)$ can only hold if $V_{L}^{2} G_{L}^{R}(\omega) + V_{R}^{2} G_{R}^{R}(\omega) \gg \Gamma^{2} G_{S}^{R}(\omega) $. In the simple case where $J_{L} = J_{R} = J$, $V_{L} = V_{R} = V$ and $T_{L} = T_{R} = T$, this imposes the condition $V \gg V^{*}(T) \equiv ( \, J \,  \Gamma^{2} T \,)^{1/4}$ or equivalently $T \ll T^{*} = V^{4} / ( \, \Gamma^{2} \, J \, )$. If $T^{*} \ll T \ll J$, we can estimate $G_{S}^{R}(\omega)$ by treating the term in $V^{2}$ as a small perturbation with respect to  the pure SYK$_{2}$ solution and we get to first order in $V^{2} / \Gamma^{2}$

\begin{equation}
    G_{S}^{R}(\omega) = - \frac{i}{\Gamma} - \frac{V_{L}^{2} G_{L}^{R}(\omega) + V_{R}^{2} G_{R}^{R}(\omega)}{2 \Gamma^{2}}, \qquad T^{*} \ll T \ll J
\label{eq:G_4_p}
\end{equation}

\section{Scaling of Thermal Conductance}

In this section we give more details on the derivation of the scaling with temperature of the thermal conductance

\begin{equation}
    \mathcal{G}(T) = \frac{N V^{2}}{2} \int \frac{d \omega}{2 \pi} \, \omega \, A_{S}^{eq}(\omega) f_{eq}(- \omega) \frac{\partial}{\partial T} \Big( A_{B}^{eq}(\omega) f_{eq}(\omega) \Big)
\label{eq:conductanceSM}
\end{equation}

discussed in the main text.\\

The simplest case is the one of the non-interacting SYK$_{2}$ reservoirs. Using the low energy expressions of the spectral densities $A_{S}^{eq}(\omega) \simeq 2 / \Tilde{\Gamma}$ and $A_{B}^{eq}(\omega) \simeq 2 / J$, the thermal conductance becomes

\begin{equation}
    \mathcal{G}_{2}(T) = \frac{N V^{2}}{\pi \Tilde{\Gamma}J} \int d \omega \,  \frac{\omega^{2}}{T^{2}} \frac{e^{\frac{\omega}{T}}}{( e^{- \frac{\omega}{T}} + 1 ) ( e^{ \frac{\omega}{T}} + 1)^{2}} = N \, \frac{\pi}{6} \frac{V^{2}}{\Tilde{\Gamma} J} \, T
\end{equation}

where we used the integral $ \int_{- \infty}^{\infty} d u \, u^{2} \, \frac{e^{u}}{( e^{- u} + 1 ) ( e^{ u} + 1)^{2}} = \frac{\pi^{2}}{6} $. One must mutiply $\mathcal{G}(T)$ by $k_{B}^{2} / \hbar$ to put back the physical dimensions which leads to the result in the main text.\\

With the strongly interacting SYK$_{4}$ reservoirs, using equation \ref{eq:G_bath} we can write

\begin{equation}
    \frac{\partial}{\partial T} \Big( A_{B}^{eq}(\omega) f_{eq}(\omega) \Big) = 2  \Big( \frac{\pi}{J^{2}} \Big)^{1/4} \frac{1}{\sqrt{2 \pi}} \frac{1}{T^{\frac{3}{2}}} \psi(u), \qquad u \equiv \frac{\omega}{T}
\end{equation}

where

\begin{equation}
    \psi(u) = - \frac{1}{2} \frac{1}{e^{u} + 1} \textrm{Re} \, \bigg( \frac{\Gamma \big(\frac{1}{4} - i \frac{u}{2 \pi} \big)}{\Gamma \big(\frac{3}{4} - i \frac{u}{2 \pi} \big)} \bigg)  - u \frac{d}{d u} \bigg[ \frac{1}{e^{u} + 1} \textrm{Re} \,   \bigg( \frac{\Gamma \big(\frac{1}{4} - i \frac{u}{2 \pi} \big)}{\Gamma \big(\frac{3}{4} - i \frac{u}{2 \pi} \big)} \bigg) \bigg]
\end{equation}

Then the thermal conductance takes the form

\begin{equation}
    \mathcal{G}_{4}(T) = \frac{N V^{2}}{2 \pi}  T\int \frac{d \omega}{T} \, \frac{\omega}{T} \frac{1}{e^{- \frac{\omega}{T}} + 1 } \psi \Big( \frac{\omega}{T} \Big) \Big( \frac{\pi}{J^{2}} \Big)^{1/4} \frac{1}{\sqrt{2 \pi T}} \, A_{S}^{eq}(\omega)
\end{equation}

If $T \gg T^{*}$, to leading order in $V^{2}$ we can replace $A_{S}^{eq}(\omega)$ in the thermal conductance by $A_{S}^{eq}(\omega) \simeq 2 / \Gamma$ and we get

\begin{equation}
    \mathcal{G}_{4}(T) = I_{4} \frac{N V^{2}}{\Gamma \sqrt{J}} \sqrt{T}, \qquad \qquad T^{*} \ll T \ll J
\end{equation}

with

\begin{equation}
    I_{4} = \Big( \frac{1}{4 \pi^{5}} \Big)^{\frac{1}{4}} \int_{- \infty}^{\infty} d u \, \frac{u}{e^{-u} + 1} \psi(u) \simeq 0.36
\end{equation}

If $T \ll T^{*}$ we must use the spectral density obtained from \ref{eq:G_S4}

\begin{equation}
    A_{S}^{eq}(\omega) = \frac{1}{V^{2}} \Big( \frac{J^{2}}{\pi} \Big)^{1/4} \sqrt{2 \pi T} \, \textrm{Re} \, \bigg( \frac{\Gamma \big(\frac{3}{4} - i \frac{\omega}{2 \pi T} \big)}{\Gamma \big(\frac{1}{4} - i \frac{\omega}{2 \pi T}} \bigg)
\end{equation}

which gives

\begin{equation}
    \mathcal{G}_{4}(T) = I_{4}^{\prime} N T, \qquad \qquad T \ll T^{*}
\end{equation}

with 

\begin{equation}
    I_{4}^{\prime} = \frac{1}{2 \pi} \int_{- \infty}^{\infty} d u \, \frac{u}{e^{- u} + 1} \, \textrm{Re} \, \bigg( \frac{\Gamma \big(\frac{3}{4} - i \frac{u}{2 \pi} \big)}{\Gamma \big(\frac{1}{4} - i \frac{u}{2 \pi} } \bigg) \psi(u) \simeq 0.16
\end{equation}

\end{document}